\documentclass[prc,preprint,showpacs,showkeys,nofootinbib]{revtex4}%
\usepackage{amssymb}
\usepackage{graphicx}
\usepackage{bm}
\usepackage{amsmath}
\usepackage{amsfonts}%
\newcommand{\beq}{\begin{equation}}
\newcommand{\eeq}{\end{equation}}
\newcommand{\bea}{\begin{eqnarray}}
\newcommand{\eea}{\end{eqnarray}}
\newcommand{\ben}{\begin{eqnarray*}}
\newcommand{\een}{\end{eqnarray*}}

\begin{document}
\title{Breakup of the weakly bound $^{17}$F nucleus}
\author{C.A. Bertulani and P. Danielewicz}
\email{bertulani@nscl.msu.edu,   danielewicz@nscl.msu.edu}
\affiliation{National Superconducting Cyclotron Laboratory,
Michigan State University, East Lansing, MI\ 48824, USA}
\date{\today }

\begin{abstract}
The breakup of the radioactive $^{17}$F nucleus into a proton and $^{16}$O is
studied for the reaction $^{17}$F$\ +\ ^{208}$Pb$\ \longrightarrow\ $p +
$^{16}$O +$\ ^{208}$Pb at 65 MeV/nucleon. The possibility of using this
reaction as a test case for studying dynamical Coulomb reacceleration effects
is assessed. It is shown that the reaction is dominated by elastic nuclear
breakup (diffraction dissociation).

\end{abstract}
\pacs{25.60Gc,24.10.-i}
\maketitle

\section{Introduction}

The Coulomb dissociation method \cite{BBR86} is now a standard
experimental tool as a source of information on radioactive
capture processes of astrophysical interest. It can be shown that
the breakup cross sections of a projectile nucleus in the Coulomb
field of a target is proportional to the cross section for
photo-dissociation \cite{BB85}. The radiative capture cross
sections can be obtained via detailed balance \cite{BBR86}. Thus,
by measuring the Coulomb breakup of nuclear projectiles, specially
at high bombarding energies ($\sim50-200$ MeV/nucleon), one can
obtain information on the radiative capture cross section of
interest. This goal has indeed been achieved in numerous
experiments for the study of $^{4}$He$\left(
\text{d},\gamma\right)  ^{6}$Li \cite{Kie93}, $^{12}$C$\left(  \text{n}%
,\gamma\right)  ^{13}$C \cite{Kie91,Mot91}, $^{11}$C$\left(  \text{p}%
,\gamma\right)  ^{12}$N \cite{Lef95}, and $^{12}$C$\left(  \alpha
,\gamma\right)  ^{16}$O \cite{Tat95}. More recently, this method has been used
exhaustively in the study of the reaction$^{7}$Be$\left(  \text{p}%
,\gamma\right)  ^{8}$B \cite{Mot94,Kik98,Iw99,Da01,Dav98,Dav01}, considered
the most important one for the standard solar model \cite{Bah98}.

Under some circumstances the relation between the Coulomb breakup
measurements and the radiative capture cross sections of interest
is not so straightforward. First the radiative capture cross
sections contain contributions of different electric and magnetic
multipolarities which enter with different weights in the Coulomb
breakup cross sections. For example, while the radiative E2
capture is very small in the reaction $^{7}$Be$\left(  \text{p}
,\gamma\right)  ^{8}$B within the solar environment, its
contribution is amplified in the Coulomb breakup experiments.
Separation of the two contributions one depends on the structure
model used for the $^{8}$B nucleus \cite{KS94,GB95,EB96, BG98}. \
Secondly, the nuclear contribution to the breakup has to be
separated from the Coulomb breakup \cite{BG98}. Finally, Coulomb
reacceleration effects \cite{BBK92,Ie93,BBe93,EBB95,NT99} have to
be controlled.

The Coulomb reacceleration effects are indeed one of the main concerns in
extracting the astrophysical S-factor for the reaction $^{7}$Be$\left(
\text{p},\gamma\right)  ^{8}$B. These effects are filtered from the data by
comparing them with dynamical calculations of Coulomb breakup
\cite{BBK92,Ie93,BBe93,EBB95,NT99}. Due to its relevance for the application
of the Coulomb dissociation method, it is desirable to study a system where the
astrophysical S-factors have been measured directly at the stellar energies
and where dynamical effects can be tested. Apparently, the breakup of $^{17}$F
is a good candidate. The ground state of $^{17}F$ is loosely-bound
(600 keV of separation
energy into proton + $^{16}$O) and its only excited state is one of the best
halo states known sofar in nuclear physics, being bound by only 100 keV.
Besides, the S-factor for the radiative capture reaction $^{16}\mathrm{O}%
\left(  \mathrm{p},\gamma\right)  ^{17}\mathrm{F}$ has been measured down to
200 keV \cite{Mor97}. Indeed, the breakup of the weakly bound $^{17}$F well
above the barrier has been considered as an important test of the dynamical
breakup mechanism \cite{Li00}. First theoretical analysis of this reaction
has been done in ref. \cite{EB02}.

In this work we study the breakup of $^{17}$F into proton + $^{16}$O, which
would be relevant for the purpose of testing Coulomb dynamical effects by a
comparison with experimental results. \ Thus, we do not consider the stripping
of the proton from $^{17}$F, or the breakup into other channels. We will only
consider two breakup mechanisms: the Coulomb and the elastic nuclear breakup
(or diffraction dissociation). If the proton and the $^{16}$O fragments are
measured in coincidence, these are the only two mechanisms of interest. We
restrict our calculation to first order perturbation theory, as we want to
learn about the feasibility of such experiments. We also restrict ourselves to
bombarding energies of 65 MeV/nucleon, typical of laboratories like
GANIL/France, GSI/Germany, NSCL/USA and RIKEN/Japan, where previous Coulomb
dissociation experiments have been carried out.

We show that the Coulomb breakup cross sections are too small to
allow for a reliable experimental counting rate. Also, the
reaction is dominated by the elastic nuclear breakup mechanism
which is more model dependent than the Coulomb breakup. The
details of our calculations are presented in section 2, where we
show that a single particle model is able to explain the
astrophysical S-factor for $^{16}\mathrm{O}\left(
\mathrm{p},\gamma\right) ^{17}\mathrm{F}$ at the lowest energies.
In section 3 we present our calculations for Coulomb breakup and in
section 4 for elastic nuclear breakup. Our conclusions are
presented in section 5.

\section{Potential model for $^{17}\mathrm{F}$}

In the present calculation, we treat $^{17}$F as the combination of
a proton and an inert $^{16}$O
core with spin 0. The proton-oxygen wavefunction is given by%
\begin{equation}
\Psi_{lj}\left(  \mathbf{r}\right)  =\frac{u_{lj}\left(  r\right)  }{r}%
\sum_{m,m_{s}}(l{\tfrac{1}{2}}m_{l}m_{s}|j{m})Y_{lm_{l}}\left(  \widehat
{\mathbf{r}}\right)  \chi_{m_{s}}\ , \label{psi_expansion}%
\end{equation}
where $\chi_{m_{s}}$\ is the spinor wavefunction.

\begin{figure}[t]
\begin{center}
\includegraphics[
height=3.4353in, width=3.4353in ]{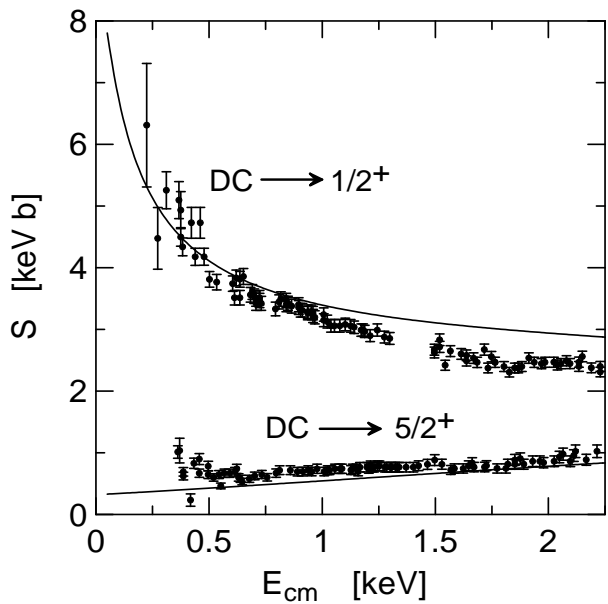}
\end{center}
\caption{S-factor for the reaction $^{16}$O$($p$,\gamma)^{17}$F.
The upper (lower) data represent the capture to the
$1/2^{+}$-excited ($5/2^{+}$-ground) state in $^{17}$F. The solid
curves are the result of a calculation for the direct capture with
a potential model described in the text. The data are from
ref. \cite{Mor97}.}%
\label{fig1}%
\end{figure}

The radial wave functions $R_{lj}(r)$ are solutions of the radial
Schr\"{o}dinger equation for the proton-core motion with the potential
\begin{equation}
V(\mathbf{r})=V_{0}\left[  1-F_{s.o.}(\mathbf{l.s})\dfrac{r_{0}}{r}\dfrac
{d}{dr}\right]  \left[  1+\exp\left(  \frac{r-R}{a}\right)  \right]
^{-1}+V_{C}(r) \label{pot_mod}%
\end{equation}
where
\begin{align}
V_{C}(r)  &  =\frac{8e^{2}}{r}\ \ \ \mathrm{\ for}\ \ \ \ \ r>R_{C}\nonumber\\
&  =\frac{4e^{2}}{r}\left(  3-\frac{r^{2}}{R_{C}^{2}}\right)
\ \ \ \ \mathrm{for}\ \ \ \ r<R_{C}. \label{coul_pot}%
\end{align}

We use $a=0.6$ fm, $R_{C}=R=3.27$ fm, $F_{s.o.}=0.341$ fm, and $r_{0}=1.25$
fm. For the $\frac{5}{2}^{+}$ d-wave\ ground state, we use the potential
depth\ $V_{0}=-49.66$ MeV, which reproduces the separation energy of 0.6 MeV.
For the $\frac{1}{2}^{+}$ \ s-wave, we use \ $V_{0}=-50.65$ MeV, which
reproduces the separation energy of 0.1 MeV of the only excited state in
$^{17}\mathrm{F}$.

The continuum wavefunctions are calculated with the same potential model
parameters as the $\tfrac{5}{2}^{+}$\ ground state. They are normalized so as
to satisfy the relation%
\begin{equation}
\left\langle \Psi_{c}|\Psi_{c^{\prime}}\right\rangle =\delta\left(
E_{c}-E_{c^{\prime}}\right)  \delta_{jj^{\prime}}\delta_{ll^{\prime}}%
\delta_{mm^{\prime},} \label{cont_norm}%
\end{equation}
what means, in practice, that the continuum wavefunctions $u_{Elj}(r)$ are
normalized to $\sqrt{2m_{bc}/\pi\hbar^{2}k}~\sin(kr+\delta_{lj})$ at large
$r$, where $k$ is the relative momentum of the fragments $b$ and $c$ (oxygen
and proton, respectively).

\begin{figure}[t]
\begin{center}
\includegraphics[
height=4.5in, width=3.in ]{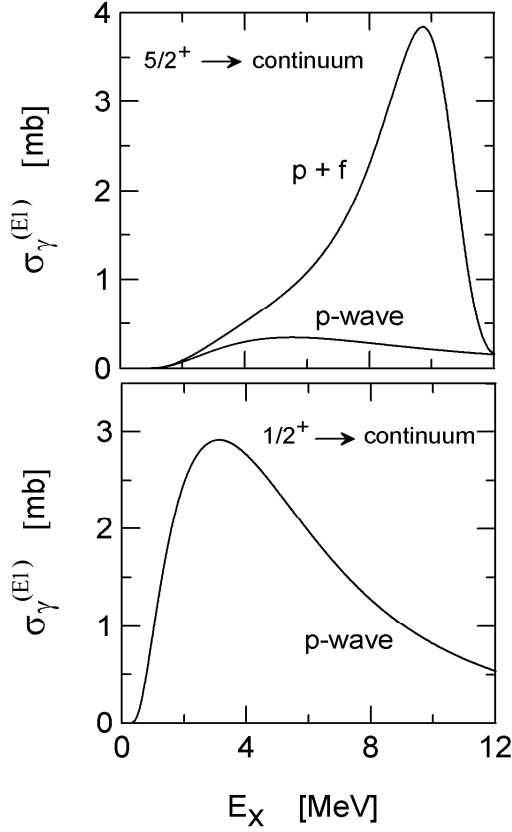}
\end{center}
\caption{The upper (lower) figure shows the photo-absorption cross
section of $^{17}\mathrm{F}$ for transitions from the ground
(first excited) state to the continuum. The curves are calculated
within the direct capture model explained
in the text and are given as a function of the photon energy $E_{\gamma}%
=E_{x}=E_{cm}+S_{p},$ where $E_{x}$ is the excitation energy and
$S_{p}$ is
the separation energy.}%
\label{fig2}%
\end{figure}

The S-factor for the direct capture from a continuum state to the bound state,
with spin $j_{0}$, is given by%
\begin{equation}
S_{\lambda}\left(  E_{cm}\right)  =\frac{\left(  2\pi\right)  ^{3}\left(
\lambda+1\right)  }{2\lambda\left[  \left(  2\lambda+1\right)  !!\right]
^{2}}\;\frac{1}{k}\;\left(  \frac{E_{x}}{\hbar c}\right)  ^{2\lambda+1}%
\exp\left[  2\pi\eta\left(  E_{cm}\right)  \right]  \ \left\vert
\mathcal{O}_{\lambda}(E_{cm};lj;l_{0}j_{0})\right\vert ^{2} \label{s_lambda}%
\end{equation}
where $\lambda=1$, or 2, for E1, or E2 transitions, respectively. The
electromagnetic matrix element $\mathcal{O}_{\lambda}(E_{cm};lj;l_{0}j_{0})$
is given by%
\begin{align}
\mathcal{O}_{\lambda}(E_{cm};lj;l_{0}j_{0})  &  =\frac{{e_{\lambda}}}%
{\sqrt{4\pi}}{\ }(-1)^{j_{0}+l_{0}+l-j}\ \left[  \frac{1+\left(  -1\right)
^{l+l_{0}+\lambda}}{2}\right]  {\frac{\hat{\lambda}\hat{j_{0}}}{\hat{\jmath}}%
}(j_{0}{\tfrac{1}{2}}\lambda0|j{\tfrac{1}{2}})\nonumber\\
&  \times\int u_{Elj}(r)\ u_{l_{0}j_{0}}(r)\ r^{\lambda}dr, \label{overlap}%
\end{align}
with the notation $\widehat{j}\equiv\sqrt{2j+1}$, and $e_{\lambda}%
=Z_{b}e(-A_{c}/A_{a})^{\lambda}+Z_{c}e(A_{b}/A_{a})^{\lambda}$ where
$a\equiv\ ^{17}$F$,$ $b\equiv\;^{16}$O and $\ c\equiv$p, respectively$.$

In a similar fashion, the photo-absorption cross section for the reaction $\gamma
+a\longrightarrow b+c$ is given by%
\begin{equation}
\sigma_{\gamma}^{(\lambda)}\left(  E_{cm}\right)  =\frac{\left(  2\pi\right)
^{3}\left(  \lambda+1\right)  }{\lambda\left[  \left(  2\lambda+1\right)
!!\right]  ^{2}}\left(  \frac{\mu_{bc}}{\hbar^{2}k}\right)  \left(
\frac{E_{x}}{\hbar c}\right)  ^{2\lambda-1}\left\vert \mathcal{O}_{\lambda
}(E_{cm};lj;l_{0}j_{0})\right\vert ^{2}, \label{sig_lambda}%
\end{equation}
while the photo-absorption cross section for the ground state to excited state
transition $\gamma+a\left(  5/2^{+}\right)  \longrightarrow a\left(
1/2^{+}\right)  $ is given by%
\begin{equation}
\sigma_{\gamma}^{(\lambda)}\left(  E_{x}\right)  =\frac{\left(  2\pi\right)
^{3}\left(  \lambda+1\right)  }{\lambda\left[  \left(  2\lambda+1\right)
!!\right]  ^{2}}\left(  \frac{E_{x}}{\hbar c}\right)  ^{2\lambda-1}\left\vert
\mathcal{O}_{\lambda}(l_{1}j_{1};l_{0}j_{0})\right\vert ^{2}\delta\left(
E_{f}-E_{i}-E_{x}\right)  , \label{phtbound}%
\end{equation}
where $\left\vert \mathcal{O}_{\lambda}(l_{1}j_{1};l_{0}j_{0})\right\vert $ is
given by eq. \ref{overlap}, with the wavefunction of the $1/2^{+}$\ state
replacing $u_{Elj}(r),$ and with $E_{f}-E_{i}=0.5$ MeV.

In figure \ref{fig1}, we show the S-factor for the radiative capture reaction
$^{16}\mathrm{O}\left(  \mathrm{p},\gamma\right)  ^{17}\mathrm{F}$. The data
points are from ref. \cite{Mor97}. \ We only show the low-energy part of the
spectrum, up to $E_{cm}=2$ MeV. \ The solid curves are the result of
calculations following the direct capture model as described above. The data
are reasonably well described by the model, but the model overestimates the capture
cross sections into the $1/2^{+}$\ state for energies greater than 1 MeV.

In figure \ref{fig2}, we show the photo-absorption cross sections for
transitions from the ground state (upper figure) and from the first excited
state (lower figure) to the continuum. The curves are calculated within the
direct capture model and are given as a function of the photon energy
$E_{\gamma}=E_{x}=E_{cm}+S_{p},$ where $E_{x}$ is the excitation energy and
$S_{p}$ is the separation energy. Only shown are the E1 transition cross
sections as the E2 cross sections are at least a factor of $10^{3}$ smaller. The
p-waves dominate the transitions for lower energies from the ground state to
the continuum, but the f-wave contributions dominate at higher energies, of 2
MeV and above. \ The photo-absorption cross section for the transitions from
the $1/2^{+}$\ to the continuum is shown in the lower panel.

\begin{figure}[t]
\begin{center}
\includegraphics[
height=3.3in, width=3.8in ]{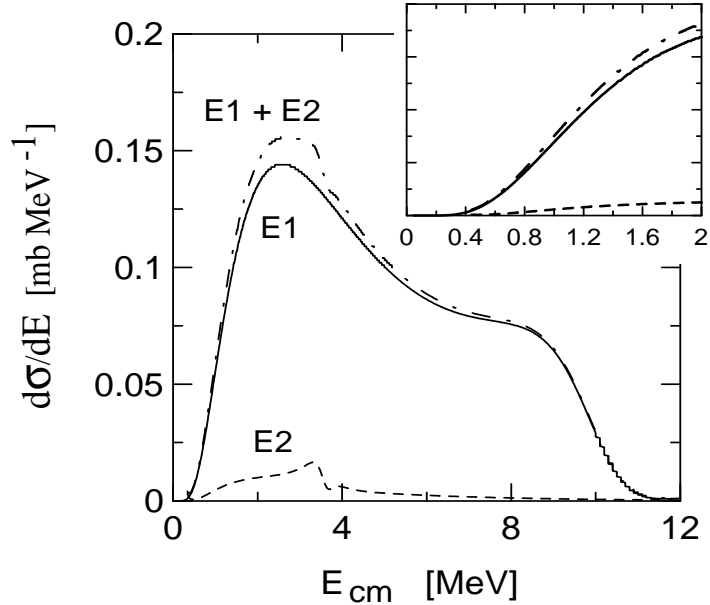}
\end{center}
\caption{Coulomb breakup cross section $d\sigma/dE$ (in mb/MeV)
for the reaction $^{17}\mathrm{F}\;$(65 MeV/nucleon)
$+\mathrm{Pb}\longrightarrow
\;^{16}\mathrm{O}+\mathrm{p}+\mathrm{Pb}$, as a function of the proton-$^{16}%
\mathrm{O}$ relative energy, in MeV. }%
\label{fig3}
\end{figure}

There is also a contribution for the excitation of the $1/2^{+}$\ bound-state
from the ground state. This contribution was calculated by using eq. \ref{phtbound}. The
cross section value integrated over the line is $\int\sigma_{\gamma}%
^{(5/2^{+}\longrightarrow1/2^{+})}\left(  E_{x}\right)  dE_{x}=7\times10^{-6}$
mb MeV. The result can be translated into a B(E2)-value for the down transition
$B(E2;1/2^{+}\longrightarrow5/2^{+})=61.5\ e^{2}$ fm$^{2}$, which is rather
close to the experimental value of $66.4\pm1.4$ e$^{2}\ $fm$^{2}$
\cite{Ajz71}. This is indeed verified in more elaborate calculations
\cite{BAM77}, the reason being that the loosely-bound states in $^{17}$F have
a small overlap with the core, thus leading to a small core polarization.
Single particle states are thus a good approximation for these states.

One observes from figure \ref{fig2} \ that the photo-absorption cross sections
have maxima at rather high energies, ($E_{\max}\sim10$ MeV for the g.s. and
$E_{\max}\sim3$ MeV for the excited state) in contrast to the photo-absorption
cross sections of neutron halo nuclei \cite{BB88a}. As explained in ref.
\cite{BB88a}, the photo dissociation cross section peaks at an energy of about
twice the binding energy of neutron halo states. The cause of the peaks
moving towards the higher energies is the Coulomb barrier which produces
a reduction of the
overlap integral in eq. \ref{overlap}. The result
is an effective separation energy that is much larger than in neutron halo
systems with the same separation energies.

\section{Coulomb breakup}

Since there are no data for the elastic scattering of $^{17}\mathrm{F}$ on
$\mathrm{Pb}$ targets at high bombarding energies, we construct an optical
potential using an effective interaction of the M3Y type \cite{KBS84,BLS97}
modified so as to reproduce the energy dependence of total reaction cross
sections, i.e. \cite{BLS97},
\begin{equation}
t(E,s)=-i{\frac{\hbar v}{2t_{0}}}\;\sigma_{NN}(E_{lab})\;\left[
1-i\alpha(E_{lab})\right]  \;t(s)\ , \label{tes}%
\end{equation}
where $t_{0}=421$ MeV fm$^{3}$ is the volume integral of the M3Y interaction
$t(s)$, $s$ is the nucleon-nucleon separation distance, $\mathrm{v}$ is the
projectile velocity, $\sigma_{NN}$ is the nucleon-nucleon cross section, and
$\alpha$ is the real-to-imaginary ratio of the forward nucleon-nucleon
scattering amplitude. At 65 MeV/nucleon, we use $\sigma_{NN}=82$ fm$^{2}$ and
$\alpha=0.96$.

The optical potential is given by
\begin{equation}
U(E_{lab},\mathbf{R})=\int d^{3}r_{1}\;d^{3}r_{2}\;\rho_{_{P}}(\mathbf{r}%
_{1})\rho_{_{T}}(\mathbf{r}_{2})\;t(E_{lab},s)\ , \label{Uopt}%
\end{equation}
where $\mathbf{s}=\mathbf{R}+\mathbf{r}_{2}-\mathbf{r}_{1}$, and $\rho_{_{T}}$
($\rho_{_{P}}$) is the ground state density of the target (projectile). For
the proton, we use a gaussian density characterized by a width of 0.7 fm. For $^{16}$O
and $^{208}$Pb we use the matter densities tabulated in ref. \cite{Vri87}.

\begin{figure}[t]
\begin{center}
\includegraphics[
height=2.in, width=3.4in ]{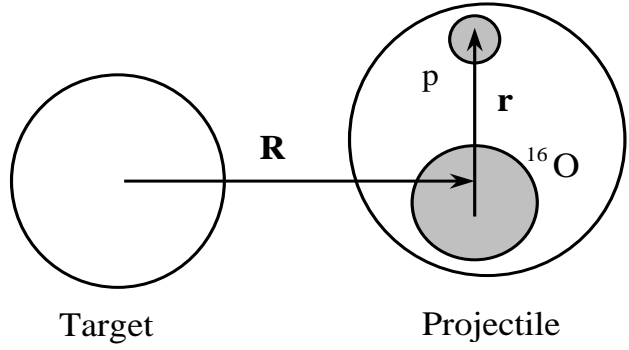}
\end{center}
\caption{Coordinates used in the text to describe the $^{17}$F breakup. }%
\label{sq}%
\end{figure}

Following ref. \cite{BN93}, the Coulomb amplitude for E1 transitions is given
by
\begin{equation}
f_{E1}^{(jm,j_{0}m_{0})}=2\sqrt{\frac{2\pi}{3}}i^{\mu}\;{\frac{Z_{T}%
e\mu_{_{PT}}}{\hbar^{2}}}\;\left(  {\frac{E_{x}}{\hbar\mathrm{v}}}\right)
\;\left\langle j_{0}m_{0}1\left(  m-m_{0}\right)  |jm\right\rangle
\ \mathcal{O}_{1}(E_{cm};lj;l_{0}j_{0})\ g_{\mu}\left(  \gamma\right)
\ \Omega_{\mu}(q), \label{f_E1}%
\end{equation}
where $\mu_{PT}$ is the reduced mass of the target + projectile, v is their
relative velocity, $\gamma=\left(  1-\mathrm{v}^{2}/c^{2}\right)  ^{-1/2}$,
and
\begin{equation}
\Omega_{\mu}(q)=\int_{0}^{\infty}db\;b\;J_{\mu}(qb)K_{\mu}\left(  {\frac
{E_{x}b}{\gamma\hbar\mathrm{v}}}\right)  \exp\left[  {i\chi(b)}\right]  \ ,
\label{omega_mu}%
\end{equation}
where $q=2p_{cm}\sin(\theta/2)$, $p_{cm}$ is the center of mass bombarding
momentum and $\theta$ is the scattering angle. The function $J_{\mu}(K_{\mu})$ is the
cylindrical (modified) Bessel function of order $\mu$. The functions $g_{\mu
}\left(  \gamma\right)  $ are given by%
\begin{equation}
g_{0}=\frac{\sqrt{2}}{\gamma},\ \ \ \ \ \ g_{1}=-g_{-1}=i\ .
\label{g_functions}%
\end{equation}

The eikonal phase, $\chi(b)$, is given by
\begin{equation}
\chi(b)=2\eta\ln(kb)-{\frac{1}{\hbar\mathrm{v}}}\;\int_{-\infty}^{\infty
}dz\;U_{opt}(R)\ , \label{eikphase}%
\end{equation}
where $\eta=Z_{P}Z_{T}e^{2}/\hbar\mathrm{v}$, $\hbar k=p_{cm}$ is the center
of mass momentum, and $R=\sqrt{b^{2}+z^{2}}$. The optical potential, $U_{opt}%
$, in the above equation is given by eq. \ref{Uopt}.

In a similar fashion, the E2 transition amplitudes are given by%
\begin{equation}
f_{E2}^{(jm,j_{0}m_{0})}=2\sqrt{\frac{\pi}{30}}i^{\mu}\;{\frac{Z_{T}%
e\mu_{_{PT}}}{\hbar^{2}}}\;\left(  {\frac{E_{x}}{\hbar\mathrm{v}}}\right)
^{2}\;\left\langle j_{0}m_{0}2\mu|jm\right\rangle \ \mathcal{O}_{2}%
(E_{cm};lj;l_{0}j_{0})\ h_{\mu}\left(  \gamma\right)  \ \Omega_{\mu}(q),
\label{fE2}%
\end{equation}
where $\mu=m-m_{0}$, and%
\begin{equation}
h_{0}=i\frac{\sqrt{6}}{\gamma},\ \ \ \ \ \ h_{1}=-h_{-1}=-\left(
1+1/\gamma^{2}\right)  ,\ \ \ \ \ \ h_{2}=h_{-2}=-i\frac{1}{\gamma}.
\label{h_functions}%
\end{equation}
The cross section for Coulomb excitation of a state with angular momentum $j$
and excitation energy $E_{x}$ is obtained by averaging (and summing) over the
initial (final) angular momentum projections:
\begin{equation}
\frac{d^{2}\sigma_{C}^{(lj)}}{d\Omega dE_{x}}={\frac{1}{2j_{0}+1}}%
\;\sum_{m_{0},\ m}\left\vert f_{E1}^{(jm,j_{0}m_{0})}+f_{E2}^{(jm,j_{0}m_{0}%
)}\right\vert ^{2}\;. \label{dsdEC}%
\end{equation}

\begin{figure}[t]
\begin{center}
\includegraphics[
height=3.in, width=3.in ]{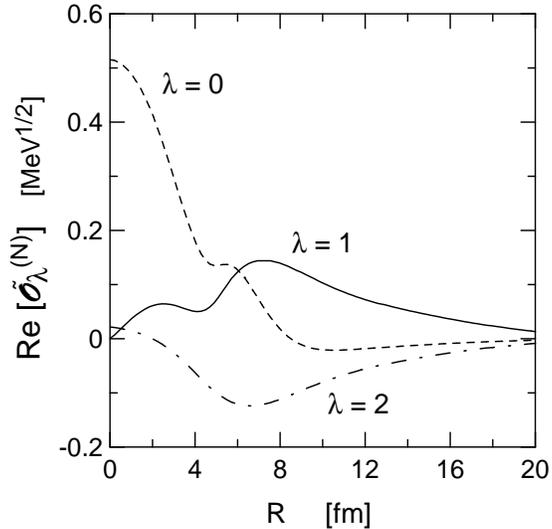}
\end{center}
\caption{ The overlap function
$\widetilde{\mathcal{O}}_{\lambda}^{(N)}$, defined by eqs.
\ref{O_dwba_1} and \ref{ovN2}, as a function of $R$, for
$\lambda=0$, 1 and 2, and for excitation energy $E_{x}=1.5$ MeV. }%
\label{ovN}%
\end{figure}

The E1 and E2 amplitudes do not interfere in the cross section after summation
over angular momentum projections. This can easily be seen using the
orthonormality condition of the Clebsh-Gordan coefficients:%
\begin{equation}
\sum_{m_{0},\ m}\left\langle j_{0}m_{0}2\mu|jm\right\rangle \left\langle
j_{0}m_{0}1\mu|jm\right\rangle =0. \label{proof_norm}%
\end{equation}
Thus, eq. \ref{dsdEC} is just a sum of cross sections for
the E1 and E2 excitations separately:
\begin{align}
\frac{d^{2}\sigma_{C}^{(lj)}}{d\Omega dE_{x}}  &  =\frac{d^{2}\sigma
_{E1}^{(lj)}}{d\Omega dE_{x}}+\frac{d^{2}\sigma_{E2}^{(lj)}}{d\Omega dE_{x}%
}\nonumber\\
&  =\frac{n_{E1}\left(  E_{x},\theta\right)  }{E_{x}}\sigma_{\gamma}^{(\lambda
j)}\left(  E_{x}\right)  +\frac{n_{E2}\left(  E_{x},\theta\right)  }{E_{x}%
}\sigma_{\gamma}^{(\lambda j)}\left(  E_{x}\right)  \ . \label{C_diss}%
\end{align}
where the virtual photon numbers $n_{E1}$ and $n_{E2}$ \ are given by%
\begin{align}
n_{E1}\left(  E_{x},\theta\right)   &  =\frac{Z_{T}^{2}\alpha}{\pi^{2}}\left(
\frac{c}{\mathrm{v}}\right)  ^{2}\left[  \zeta_{1}\left(  \omega\right)
+\frac{1}{\gamma^{2}}\zeta_{0}\left(  \omega\right)  \right] \nonumber\\
n_{E2}\left(  E_{x},\theta\right)   &  =\frac{Z_{T}^{2}\alpha}{\pi^{2}}\left(
\frac{c}{\mathrm{v}}\right)  ^{4}\left[  \frac{1}{\gamma^{2}}\zeta_{2}\left(
\omega\right)  +\left(  1+\frac{1}{\gamma^{2}}\right)  ^{2}\zeta_{1}\left(
\omega\right)  +\frac{3}{\gamma^{2}}\zeta_{0}\left(  \omega\right)  \right]
\ , \label{virt}%
\end{align}
with $\alpha=1/137$ and%
\begin{equation}
\zeta_{\mu}\left(  \omega\right)  =2\pi\left(  \frac{\omega}{\gamma\mathrm{v}%
}\right)  ^{2}\int db\ b\ K_{\mu}^{2}\left(  \frac{\omega b}{\gamma\mathrm{v}%
}\right)  \exp\left[  -2\operatorname{Im}\chi\left(  b\right)  \right]  \ .
\label{zeta}%
\end{equation}

Figure 3 shows the Coulomb breakup cross section $d\sigma/dE$ (in mb/MeV) for
the reaction $^{17}\mathrm{F}\;$(65 MeV/nucleon) $+\ \mathrm{Pb}%
\longrightarrow\;^{16}\mathrm{O}+\mathrm{p}+\mathrm{Pb}$, as a function of the
proton-$^{16}\mathrm{O}$ relative energy, in MeV. The smaller panel on the
upper right corner of the figure shows the cross section at energies up to
$E_{cm}=2$ MeV. One sees that the E1 breakup mode is dominant at all energies.

Although the E2 contribution to the photo-dissociation cross section is very
small it is amplified in the Coulomb dissociation due to the large abundance of E2
photons at the bombarding energy considered \cite{BB88}. However, the E2
contribution to $d\sigma/dE$ \ is only relevant at $E_{cm}\sim3$ MeV where the
cross section has a peak. We also note that this cross section is about a
factor $10^{3}$ smaller than the Coulomb breakup cross section for the
reaction $^{8}\mathrm{B}\;$ $+\ \mathrm{Pb}\longrightarrow\;^{7}%
\mathrm{Be}+\mathrm{p}+\mathrm{Pb}$ at similar bombarding energies
\cite{Mot94,Kik98,Iw99,Da01,Dav98,Dav01}. Thus, it is much more likely that
the breakup is dominated by the nuclear interaction in this case. We will demonstrate
this to be indeed the case in the next section.

\begin{figure}[t]
\begin{center}
\includegraphics[
height=3.in, width=3.in ]{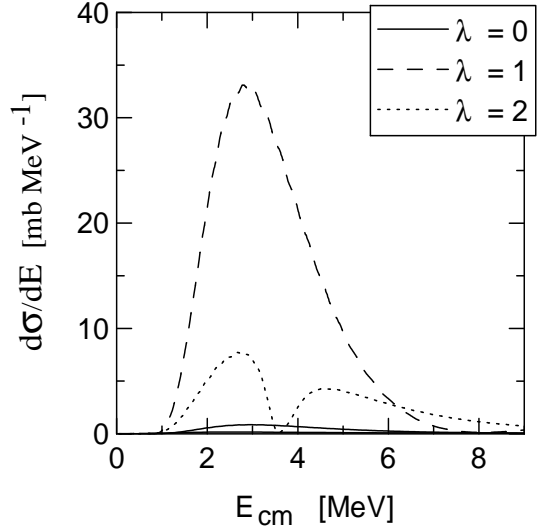}
\end{center}
\caption{Elastic nuclear breakup cross section $d\sigma/dE$ (in
mb/MeV) for
the reaction $^{17}\mathrm{F}\;$(65 MeV/nucleon) $+\ \mathrm{Pb}%
\longrightarrow\;^{16}\mathrm{O}+\mathrm{p}+\mathrm{Pb}$, as a
function of the
proton-$^{16}\mathrm{O}$ relative energy, in MeV. }%
\label{dsde_N}%
\end{figure}

The angle integrated cross sections for ground state to continuum transitions
are $\sigma_{E1}=0.88$ mb, $\sigma_{E2}=0.054$ mb and $\sigma_{\mathrm{total}%
}=0.94$ mb, respectively. This should be compared to the cross section for the
excitation of the 1/2$^{+}$ state which is 0.33 mb. These cross sections are
small and hard to measure with reliable accuracy at present laboratory
facilities. Also, since the strength of the Coulomb breakup is much reduced
for $^{17}\mathrm{F}$ as compared to the breakup of $^{8}\mathrm{B}$
projectiles, one also expects that higher order effects, which play a role in
the breakup of $^{8}\mathrm{B}$, become smaller in the case of
$^{17}\mathrm{F}$ breakup.

\section{Nuclear breakup}

The nuclear breakup is due to the nuclear interaction
\begin{equation}
U\left(  \mathbf{R,r}\right)  =U_{\mathrm{p}}\left(  \mathbf{r}_{\mathrm{p}%
}\right)  +U_{\mathrm{O}}\left(  \mathbf{r}_{\mathrm{O}}\right)
-U_{\mathrm{F}}\left(  \mathbf{R}\right)  , \label{U_opt_3}%
\end{equation}
where $U_{\mathrm{p}}\left(  \mathbf{r}_{\mathrm{p}}\right)  $, $U_{\mathrm{O}%
}\left(  \mathbf{r}_{\mathrm{O}}\right)  ,$ and $U_{\mathrm{F}}\left(
\mathbf{R}\right)  $ are the optical potentials for the scattering of the
proton, oxygen, and fluorine, off the target, respectively. The vectors
$\mathbf{r}_{\mathrm{p}}$,\ $\mathbf{r}_{\mathrm{O}}$, and $\mathbf{R}$ are
their respective coordinates relative to the target (see figure \ref{sq}%
).\ The fluorine-projectile potential is only responsible for the center of
mass scattering and does not influence the $^{17}$F breakup.

The DWBA matrix element for the breakup,
under the assumption of spherically symmetric  potentials, is given by%
\begin{equation}
T_{\mathrm{N}}=\left\langle \Phi^{\left(  -\right)  }\left(  \mathbf{R}%
\right)  \Psi_{E}\left(  \mathbf{r}\right)  \left\vert U_{\mathrm{p}}\left(
\left\vert \mathbf{R}-\beta_{1}\mathbf{r}\right\vert \right)  +U_{\mathrm{O}%
}\left(  \left\vert \mathbf{R}-\beta_{2}\mathbf{r}\right\vert
\right) \right\vert \Psi_{g.s.}\left(  \mathbf{r}\right)
\Phi^{\left(  +\right)
}\left(  \mathbf{R}\right)  \right\rangle , \label{nuc_DWBA}%
\end{equation}
where \textbf{r} denotes the relative coordinate between the
proton and the oxygen core in $^{17}\mathrm{F}$, $\Phi^{\left(
-\right)  }\left(  \mathbf{R}\right)  $ is the ingoing and
$\Phi^{\left(  +\right)  }\left(  \mathbf{R}\right)  $ is the
outgoing center of mass (c.m.) scattering wavefunction, and where
$\beta_{1}=16/17$ and $\beta_{2}=-1/17$ (see fig. \ref{sq}).

\begin{figure}[t]
\begin{center}
\includegraphics[
height=3.in, width=3.in ]{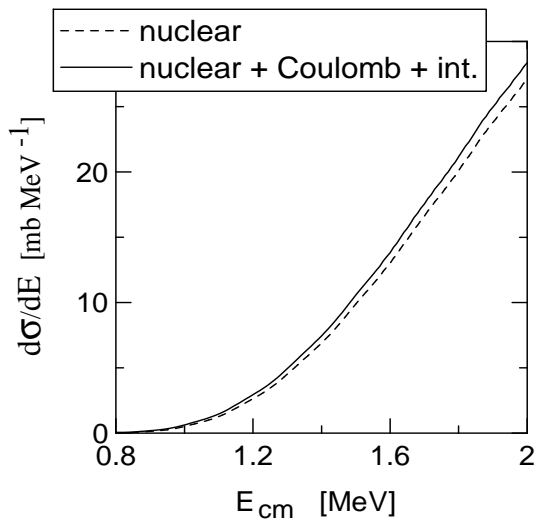}
\end{center}
\caption{Elastic nuclear breakup (dashed line) cross section
$d\sigma/dE$ (in
mb/MeV) for the reaction $^{17}\mathrm{F}\;$(65 MeV/nucleon) $+\ \mathrm{Pb}%
\longrightarrow\;^{16}\mathrm{O}+\mathrm{p}+\mathrm{Pb}$, as a
function of the proton-$^{16}\mathrm{O}$ relative energy, in MeV.
The solid line includes the Coulomb breakup and the
nuclear-Coulomb interference.} \label{cnint}
\end{figure}

For the c.m. scattering at high energies one can use the eikonal
approximation which implies
\begin{equation}
\Phi^{\left(  -\right)  \ast}\left(  \mathbf{R}\right)
\Phi^{\left(
+\right)  }\left(  \mathbf{R}\right)  =\exp\left[  -i\mathbf{q\cdot R}%
+i\chi(b)\right] \ , \label{eikonal_psi}%
\end{equation}
where $\mathbf{q=k}^{\prime}\mathbf{-k}$\ is the momentum transfer
to the c.m. and $\chi(b)$\ is the eikonal phase for the c.m.
scattering, given by eq. \ref{eikphase}. Since the momentum
transfer is small compared to the total projectile momentum, we
can use $q=2p_{cm}\sin(\theta/2)$, where $\theta$ is the
scattering angle.\

It is appropriate to expand the potentials $U_{i}\left( \left\vert
\mathbf{R}-\beta_{i}\mathbf{r}\right\vert \right)  $ into
multipoles to exploit the spherical symmetry of the states in
$^{17}\mathrm{F}$. One has then
\begin{equation}
U_{i}\left(  \left\vert \mathbf{R}-\beta_{i}\mathbf{r}\right\vert
\right) =\sum_{\lambda}\mathcal{F}_{\lambda}^{(i)}\left(
r,R\right)  P_{\lambda }\left(  \cos\theta\right)
=\sum_{\lambda\mu}\frac{4\pi}{2\lambda
+1}\mathcal{F}_{\lambda}^{(i)}\left(  r,R\right)  Y_{\lambda\mu}%
(\widehat{\mathbf{R}})Y_{\lambda\mu}^{\ast}(\widehat{\mathbf{r}})\
,
\label{U_multi}%
\end{equation}
where
\begin{equation}
\mathcal{F}_{\lambda}^{(i)}\left(  r,R\right)
=\frac{2\lambda+1}{2}\int
_{-1}^{1}d\xi\ P_{\lambda}\left(  \xi\right)  U_{i}\left(  \sqrt{\beta_{i}%
^{2}r^{2}+R^{2}-2\beta_{i}rR\xi}\right)  \ , \label{form_nuc}%
\end{equation}
and $P_{\lambda}\left(  \xi\right)  $ are the Legendre
polynomials.

Employing eq. \ref{psi_expansion} and the angular momentum
algebra, we get for the case of excitation into a continuum state
with angular momentum quantum numbers $jm$
\begin{align}
T_{\mathrm{N}}^{(j_{0}m_{0};jm;\lambda)}  &  =-i\sqrt{4\pi}(-1)^{-m_{0}%
}\left\langle j_{0}m_{0}\lambda\mu|jm\right\rangle \left[
\frac{1+\left(
-1\right)  ^{l+l_{0}+\lambda}}{2}\right]  {\frac{\hat{j_{0}}\hat{\jmath}}%
{\hat{\lambda}^{3}}}(j_{0}{\tfrac{1}{2}}\lambda0|j{\tfrac{1}{2}})\nonumber\\
&  \times\ \int_{0}^{\infty}dr\ u_{Elj}(r)u_{l_{0}j_{0}}(r)\ \mathcal{G}%
_{\lambda\mu}\left(  r,\mathbf{q}\right)  , \label{T_dwba_1}%
\end{align}
where $\mu=m-m_{0}$, the function ${\cal G}$ is
\begin{align}
\mathcal{G}_{\lambda\mu}\left(  r,\mathbf{q}\right)   &  =\int d^{3}%
R\ \exp\left[  -i\mathbf{q\cdot R}+i\chi(b)\right]
\mathcal{F}_{\lambda
}\left(  r,R\right)  Y_{\lambda\mu}^{\ast}(\widehat{\mathbf{R}})\nonumber\\
&  =4\pi
i^{\lambda}Y_{\lambda\mu}^{\ast}(\widehat{\mathbf{q}})\int
_{0}^{\infty}dR\ R^{2}\ j_{\lambda}\left(  qR\right)  \
\mathcal{F}_{\lambda
}\left(  r,R\right)  \ \exp\left[  i\chi(R)\right]  \ , \label{glmu}%
\end{align}
and $\mathcal{F}_{\lambda}\left(  r,R\right)
=\sum_{i}\mathcal{F}_{\lambda }^{(i)}\left(  r,R\right)  $.

In obtaining the last result in the equation above we have used
the approximation $\chi(b)\sim\chi(R)$ which is here well
justified, since the breakup occurs at the contact surface between
the two nuclei, at which point $z\sim0$. Another simplifying
approximation is to assume that the momentum transfer to the c.m.
is mostly in the perpendicular direction, following the spirit of
the eikonal
approximation. Thus, we make the substitution $Y_{\lambda\mu}^{\ast}(\widehat{\mathbf{q}%
})\longrightarrow Y_{\lambda\mu}^{\ast}(\frac{\pi}{2})$, and eq.
\ref{glmu}\ becomes
\begin{align}
\mathcal{G}_{\lambda\mu}\left(  r,\mathbf{q}\right)   &
=i^{\lambda}\left( -1\right)  ^{\left(  \lambda+\mu\right)
/2}\left[  \frac{1+\left(  -1\right)
^{\lambda+\mu}}{2}\right]  \left[  \frac{1+\left(  -1\right)  ^{l+l_{0}%
+\lambda}}{2}\right]  \frac{\sqrt{4\pi\left(  2\lambda+1\right)
\left( \lambda-\mu\right)  !\left(  \lambda+\mu\right)  !}}{\left(
\lambda
-\mu\right)  !!\left(  \lambda+\mu\right)  !!}\nonumber\\
&  \times\int_{0}^{\infty}dR\ R^{2}\ j_{\lambda}\left(  qR\right)
\ \mathcal{F}_{\lambda}\left(  r,R\right)  \ \exp\left[
i\chi(R)\right]  \ . \label{glmu_2}
\end{align}

\begin{figure}[t]
\begin{center}
\includegraphics[
height=3.in, width=3.in ]{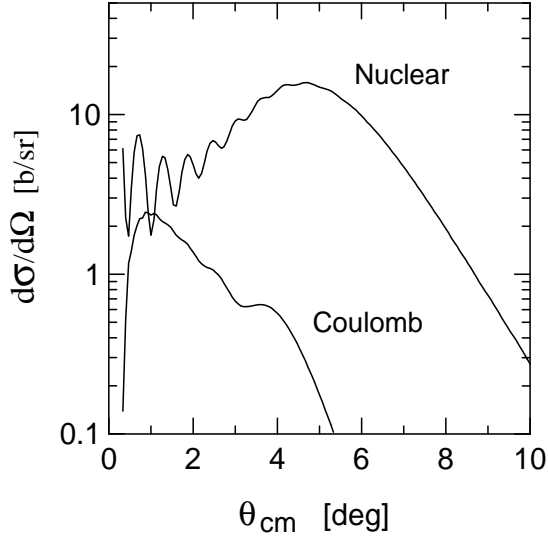}
\end{center}
\caption{Elastic nuclear breakup (upper curve) cross section
$d\sigma/d\Omega$ (in
b/sr) for the reaction $^{17}\mathrm{F}\;$(65 MeV/nucleon) $+\ \mathrm{Pb}%
\longrightarrow\;^{16}\mathrm{O}+\mathrm{p}+\mathrm{Pb}$, as a
function of the proton + $^{16}\mathrm{O}$ center of mass
scattering angle, in degrees. The lower curve line represents the
Coulomb
breakup contribution for the same reaction.}%
\label{CN_comp}%
\end{figure}

The (elastic) nuclear breakup cross scattering amplitude is given
by ($\lambda+\mu=\mathrm{even}$)
\begin{align}
&  f_{\mathrm{N}}^{(j_{0}m_{0};jm;\lambda)}\left(  \theta\right)
=-{\frac
{\mu_{_{PT}}}{2\pi\hbar^{2}}}T_{\mathrm{N}}^{(j_{0}m_{0};jm;\lambda
)}\nonumber\\
&  =i{\frac{2\mu_{_{PT}}}{\hbar^{2}}\ }(-1)^{-m_{0}+(\lambda+\mu
)/2}\ \left\langle j_{0}m_{0}\lambda\mu|jm\right\rangle \left[
\frac {1+\left(  -1\right)  ^{\lambda+\mu}}{2}\right]
\frac{\sqrt{\left( \lambda-\mu\right)  !\left(  \lambda+\mu\right)
!}}{\left(  \lambda
-\mu\right)  !!\left(  \lambda+\mu\right)  !!}\nonumber\\
&  \times\int dR\ R^{2}\ j_{\lambda}\left(  qR\right)  \exp\left[
i\chi(R)\right]  \
\mathcal{O}_{\lambda}^{(N)}(E_{cm};R;lj;l_{0}j_{0})\ ,
\label{fDWBA}
\end{align}
where
\begin{equation}
\mathcal{O}_{\lambda}^{(N)}(E_{cm};R;lj;l_{0}j_{0})=\left[
\frac{1+\left(
-1\right)  ^{l+l_{0}+\lambda}}{2}\right]  \ {\frac{\hat{j_{0}}\hat{\jmath}%
}{\hat{\lambda}^{2}}}(j_{0}{\tfrac{1}{2}}\lambda0|j{\tfrac{1}{2}})\
\int _{0}^{\infty}dr\ u_{Elj}(r)\ u_{l_{0}j_{0}}(r)\
\mathcal{F}_{\lambda}\left( r,R\right)  \ , \label{O_dwba_1}
\end{equation}
and the cross section is
\begin{align}
\frac{d\sigma_{N}^{(lj)}}{d\Omega dE}  & ={\frac{1}{2j_{0}+1}}\sum
_{m_{0},\ m,\ \lambda}\left\vert
f_{\mathrm{N}}^{(j_{0}m_{0};jm;\lambda
)}\left(  \theta\right)  \right\vert ^{2}\nonumber\\
&  =\left(  {\frac{2\mu_{_{PT}}}{\hbar^{2}}}\right)  ^{2}\sum_{\
\lambda }C\left(  \lambda jj_{0}\right)  \ \left\vert \int dR\
R^{2}\ j_{\lambda
}\left(  qR\right)  \exp\left[  i\chi(R)\right]  \mathcal{O}_{\lambda}%
^{(N)}(E_{cm};R;lj;l_{0}j_{0})\right\vert ^{2}\ \nonumber\\
&  =\left(  {\frac{2\mu_{_{PT}}}{\hbar^{2}}}\right)  ^{2}\sum_{\
\lambda }\left\vert \int dR\ R^{2}\ j_{\lambda}\left(  qR\right)
\exp\left[
i\chi(R)\right]  \widetilde{\mathcal{O}}_{\lambda}^{(N)}(E_{cm};R;lj;l_{0}%
j_{0})\right\vert ^{2}, \label{dsdO_N}%
\end{align}
with
\[
C\left(  \lambda jj_{0}\right)  ={\frac{1}{2j_{0}+1}}\sum_{\substack{m_{0}%
,\ m\\(\lambda+\mu=\mathrm{even})}}\left\langle j_{0}m_{0}\lambda
\mu|jm\right\rangle ^{2}\frac{\left(  \lambda-\mu\right)  !\left(
\lambda +\mu\right)  !}{\left[  \left(  \lambda-\mu\right)
!!\left(  \lambda +\mu\right)  !!\right]  ^{2}}\ ,
\]
and
\begin{equation}
\widetilde{\mathcal{O}}_{\lambda}^{(N)}(E_{cm};R;lj;l_{0}j_{0})=\sqrt{C\left(
\lambda jj_{0}\right)  }\ \mathcal{O}_{\lambda}^{(N)}(E_{cm};R;lj;l_{0}%
j_{0})\ . \label{ovN2}%
\end{equation}

In figure \ref{ovN}, we plot the real part of the overlap function
$\widetilde{\mathcal{O}}_{\lambda}^{(N)}$\ as a function of $R$,
for $\lambda=0$, 1 and 2, at $E_{cm}=1$ MeV. For $\lambda=0$, only
the transition from the d$\frac{5}{2}$ ground state to the
d$\frac{5}{2}$ continuum states is accounted for. For $\lambda=1$,
transitions to the p$\frac{3}{2}$, f$\frac {5}{2}$, and
f$\frac{7}{2}$ are included, while for $\lambda=2$ the transitions
to the s$\frac{1}{2}$, d$\frac{3}{2}$ and d$\frac{5}{2}$ were
considered. One observes that the overlap function $\widetilde{\mathcal{O}%
}_{\lambda}^{(N)}$\ extends to large distances $R$ between the
c.m. of the two nuclei (especially for $\lambda=1,\ 2$). The low
distance part of this function is not relevant, due to the nuclear
absorption at low impact parameters (included in the factor
$\exp\left[  i\chi(R)\right]  $ of eq. \ref{dsdO_N}). The
imaginary part of $\widetilde{\mathcal{O}}_{\lambda}^{(N)}$ has a
similar behavior as the real part.

In figure \ref{dsde_N}, we plot the elastic nuclear breakup \
cross section as a function of energy, obtained by integrating eq.
\ref{dsdO_N} over angles. The dashed, solid and dashed-dotted
curves represent the contributions of the $\lambda=0$, 1, and 2
multipolaritites, respectively. One observes that the nuclear
breakup cross sections are at least two orders of magnitude larger
than the Coulomb cross sections, as diplayed in figure \ref{fig3}.
The elastic nuclear breakup also leads to fragments with larger
relative energies. We have noticed this result does not vary
appreciably with the bombarding energy.

In figure \ref{cnint}, we plot the total (solid curve) breakup
cross section as a function of the relative energy of the proton +
$^{16}$O. The cross section includes elastic nuclear and Coulomb
breakup and Coulomb-nuclear interference. The figure is shown for
the energies of interest for astrophysics, up to 2 MeV. The dashed
curve is for the elastic nuclear breakup only. One clearly sees
that the Coulomb breakup accounts only for a tiny fraction of the
total breakup cross section.

Finally, in figure \ref{CN_comp} the angular distribution of the
Coulomb and nuclear breakup modes are compared. The distribution
was obtained by integration of the double differential cross
section, eq. \ref{dsdO_N}, over energy. One observes that even at
the very forward angles where the Coulomb excitation is by
comparison strongest, the nuclear breakup dominate.

\section{Conclusions}

We have analyzed the possibility of using the reaction
$^{17}\mathrm{F}\;$(65
MeV/nucleon) $+\ \mathrm{Pb}\longrightarrow\;^{16}\mathrm{O}+\mathrm{p}%
+\mathrm{Pb}$ for the purpose of studying the Coulomb
reacceleration effects.

We have shown that the exclusive cross section is dominated by the
elastic nuclear breakup. This shows that this reaction is not very
useful for this purpose, even if the impact parameter is selected
by imposing an upper cut on the center of mass scattering angle.
In this case, the contribution of the nuclear dissociation is
reduced considerably, as shown in ref. \cite{EB02}. However, the
Coulomb breakup cross section is also reduced. \ Since the total
Coulomb breakup cross section only amounts to fractions of
milibarns, we predict a very small counting rate for an experiment
dedicated to this study.

\section*{Acknowledgements}

This research was supported in part by the U.S. National Science
Foundation under Grants No. PHY-007091 and PHY-00-70818.

\end{document}